\documentclass[aps,amsmath]{revtex4}
\usepackage{graphics,epsf,epsfig}
\usepackage{bm}

\begin{document}

\title{Tilt grain boundary instabilities in three dimensional lamellar
  patterns}
\author{Zhi-Feng Huang and Jorge Vi\~nals}
\affiliation{McGill Institute for Advanced Materials and Department of 
Physics, McGill University, Montreal, QC H3A 2T8, Canada}

\date{\today; to be published in Phys. Rev. E}

\begin{abstract}
We identify a finite wavenumber instability of a 90$^{\circ}$ tilt grain
boundary in three dimensional lamellar phases which is absent in two dimensional
configurations. Both a stability analysis of the slowly varying amplitude or
envelope equation for the boundary, and a direct numerical solution of an
order parameter model equation are presented. The instability mode involves
two dimensional perturbations of the planar base boundary, and is suppressed
for purely one dimensional perturbations. We find that both the
most unstable wavenumbers and their growth rate increase with $\epsilon$, the
dimensionless distance away from threshold of the lamellar phase.
\end{abstract}

\maketitle

\section{Introduction}
\label{sec:introduction}

The longest relaxation times in partially ordered structures outside of
thermodynamic equilibrium are often determined by existing topological
defects. Therefore defect structure, interaction, and motion typically
control the evolution toward an equilibrium, spatially ordered state.
Prototypical example systems that are the subject of current research include
Rayleigh-B\'{e}nard convection above onset 
\cite{re:cross93,re:manneville90}, and microphase separation in block copolymers 
\cite{re:fredrickson96}. Whereas the former is an effectively two dimensional
system (fluctuations in the heat transport direction are unimportant near the
convective threshold), the latter can involve both two dimensional systems
(thin films) as well as three dimensional bulk samples. In both cases,
a transient macroscopic sample usually consists of differently oriented domains
(or grains) of the ordered phase with a large number of defects like grain 
boundaries, dislocations, and disclinations. In order to understand the
evolution of such polycrystalline state, extensive 
efforts, both theoretical and experimental, have been devoted to the
motion of these topological defects, as well as to the related
issue of wavenumber adjustment and selection.

Of particular interest is recent research on the dynamics of grain
boundaries in roll (Rayleigh-B\'{e}nard convection) or lamellar
(block copolymer) configurations. Most theoretical studies 
\cite{re:manneville83b,re:tesauro87,re:malomed90,re:boyer01b,re:huang03,%
re:huang04} are based on phenomenological order parameter models or on
amplitude equations which focus on slowly evolving amplitudes
of base patterns close to threshold. In two dimensions, a stationary
90$^{\circ}$ tilt grain boundary configuration has been found that has the same 
characteristic wavenumber ($q_0$) in both ordered domains on either
side of the boundary \cite{re:manneville83b,re:tesauro87}.
Even though for an infinite roll (or lamellar) configuration there exists 
a finite range of possible wavenumber values (i.e., the ``stability
balloon'' \cite{re:busse78,re:greenside84,re:cross93} in the two dimensional
stability diagram), phase-winding motion \cite{re:tesauro87} yields a
wavenumber selection mechanism for the grain boundary configuration, leading to 
a unique value of $q_0$ for both bulk domains. When the wavenumbers on either 
side of the boundary are different \cite{re:manneville83b}, or when both
differ from $q_0$ \cite{re:tesauro87}, boundary motion follows. Boundary
motion can be further induced by roll curvature \cite{re:boyer01b}, or by an 
externally imposed shear flow \cite{re:huang03,re:huang04}. 

Most of the known theoretical or numerical results concern two dimensional
configurations. In this paper, on the other hand, we focus
on a three dimensional configuration that also involves a $90^{\circ}$ 
tilt grain boundary. Contrary to the two dimensional result, this grain
boundary configuration is found to be always unstable. The characteristic
wavevector for instability is two dimensional (purely one dimensional
perturbations are stable), with both wavenumber components smaller than that
of the base lamellar pattern ($q_0$), and decreasing as the system approaches the
order-disorder threshold. The amplitude equation analysis reveals that
the three dimensional instability is associated with a nonzero imaginary part 
(or phase) of the complex amplitude which grows in time around the boundary
region. This is contrast with perturbations of the two dimensional stationary 
state (with a selected wavenumber $q_0$) which can be described by real
amplitudes alone \cite{re:tesauro87,re:manneville83b,re:manneville90}. The
results of the asymptotic analysis have been verified by direct numerical
solution of the Swift-Hohenberg model of convection, and good agreement is
found between the power spectra of boundary perturbations numerically
determined and the results of the stability analysis based on the amplitude
equations. Given that the model equations are of gradient or potential form,
we compute the temporal evolution of the energy following the
instability. We show that cross roll coupling terms enhance instability in
both two and three dimensions. However, these terms remain sufficiently small
in the two dimensional case and are always canceled by larger, stabilizing
terms. In three dimensions, the larger phase space available for perturbation
leads to
the instability. From these results, we would conclude that 90$^{\circ}$ tilt
grain boundaries should be rarely observed in three dimensional systems, and
even then only as transients. There is some evidence in block copolymers that
this is in fact the case \cite{re:gido94}. 

In an effort to find regions of stability, we have investigated a number of 
configurations involving a range of wavenumbers, but we have not
found any that ultimately leads to a stationary state, perhaps with a
distorted boundary. Therefore we cannot comment at this point on
the nonlinear growth and possible saturation of the instability, or on whether 
unstable motion will eventually lead to boundary annihilation. 

Section \ref{sec:stability} describes the geometry of the grain boundary 
configuration studied, and a weakly nonlinear analysis leading to the
stability calculation. The ranges and modes of instability are identified, and
the results compared with direct numerical solution of the Swift-Hohenberg
model equation. In Sec. \ref{sec:cross-roll} we examine possible instability
mechanisms for this case through the comparison between two dimensional and
three dimensional configurations. Finally, the conclusions following from
our results are summarized in Sec. \ref{sec:discussion}.

\section{Stability analysis of a 90$^{\circ}$ tilt grain boundary
  in three dimensions}
\label{sec:stability}

Our analysis is based on the Swift-Hohenberg
model equation \cite{re:swift77}, an order parameter equation originally 
developed to study the convective instability in Rayleigh-B\'enard convection, 
and later used to describe lamellae formation and reorientation in 
block copolymer melts \cite{re:fredrickson94}. In dimensionless units, the 
equation is
\begin{equation}
\frac{\partial \psi }{\partial t} 
= \epsilon \psi - (\nabla^2 + q_0^2)^2 \psi - \psi^3,
\label{eq_s-h}
\end{equation}
where for diblock copolymers the order parameter field $\psi$
represents the local density difference between two constituent
monomers, and $\epsilon$ is the dimensionless distance from
the order-disorder threshold. For $\epsilon >0$ there exists a range of
periodic stationary solutions around the wavenumber $q_{0}$ that are linearly 
stable. In the dimensionless units used
$q_{0} = 1$. We retain the symbol $q_{0}$ in what follows for the sake of
clarity in the presentation.

The system configuration studied here comprises a planar grain boundary
separating two perfectly ordered, but differently oriented lamellar
domains in three dimensions. We focus on the case of
$90^{\circ}$ grain boundary, as presented in Fig. \ref{fig_3d}, for
which the lamellar orientations of two domains A and B are
perpendicular to each other: $q_A=q_{0} \hat{x}$ for domain A, and 
$q_B=q_0 \hat{z}$ for domain B. It is well known that in two dimensions
(a grain boundary line) \cite{re:cross93,%
re:manneville83b,re:tesauro87} this choice of
wavenumber leads to a stable grain boundary configuration
against any small perturbation; however, this is not the case for 3D
system, as will be shown below.

We introduce a standard multiple scale expansion of Eq. (\ref{eq_s-h}) 
to derive the associated amplitude equations for the $90^\circ$ tilt
grain boundary. The order parameter field $\psi$ is 
expanded in both regions A and B as the superposition of two base
modes $e^{iq_0 x}$ and $e^{iq_0 z}$,
\begin{equation}
\psi=\frac{1}{\sqrt{3}} \left ( A e^{iq_0 x} 
+ B e^{iq_0 z} +{\rm c.c.} \right ),
\label{eq_expan}
\end{equation}
with complex amplitudes $A$ and $B$ that are slowly varying in space and
time. We set $T=\epsilon t$ for slow time scale, and introduce
anisotropic slow spatial scalings, with $X=\epsilon^{1/2}x,
Y=\epsilon^{1/4}y, Z=\epsilon^{1/4}z$ for mode $e^{iq_0 x}$, and
$\bar{X}=\epsilon^{1/4}x, \bar{Y}=\epsilon^{1/4}y, 
\bar{Z}=\epsilon^{1/2}z$ for mode $e^{iq_0 z}$. Then the governing
equations for amplitudes $A$ and $B$ can be obtained (to
${\cal O}(\epsilon^{3/2})$)
\begin{eqnarray}
\partial_t A &=& \left [\epsilon- ( 2 i q_0 \partial_{x} 
+\partial_y^2 +\partial_z^2 )^2 \right ] A -|A|^2 A -2 |B|^2 A, 
\label{eq_A} \\
\partial_t B &=& \left [\epsilon- ( \partial^2_{x}+ \partial^2_{y}
+2 i q_0 \partial_{z} )^2 \right ] B -|B|^2 B -2 |A|^2 B,
\label{eq_B}
\end{eqnarray}
where we have reintroduced the original spatial and temporal variables.
We note that the only difference with the amplitude equations for a two
dimensional system as given in Ref. \onlinecite{re:tesauro87} is the extra
term proportional to $\partial_y^2$ in each equation, a term which is
trivially due to the additional spatial direction available in three
dimensional space.

We first construct a base state solution involving a stationary and planar
grain boundary. Since the boundary lies on the $yz$ plane,
the amplitudes are only a function of the normal coordinate $x$:
\begin{eqnarray}
&&\epsilon A_0 + 4q_0^2 \partial_x^2 A_0 -|A_0|^2 A_0 -2 |B_0|^2 A_0=0,
\label{eq_A0} \\
&&\epsilon B_0 -\partial_x^4 B_0 -|B_0|^2 B_0 -2 |A_0|^2 B_0=0.
\label{eq_B0}
\end{eqnarray}
Although the nontrivial stationary solution $A_{0}(x)\neq 0, B_{0}(x) \neq 0$
cannot be obtained in closed form, its properties have been extensively
studied \cite{re:manneville83b,re:tesauro87}.

The complex amplitudes $A$ or $B$ are next expanded in Fourier series as
\begin{eqnarray}
A(x,y,z,t) & = & A_0(x) + \sum\limits_{q_y,q_z} \hat{A}(q_y,q_z,x,t) 
e^{i(q_y y + q_z z)}, \label{eq_Aexpan} \\
B(x,y,z,t) & = & B_0(x) + \sum\limits_{q_y,q_z} \hat{B}(q_y,q_z,x,t) 
e^{i(q_y y + q_z z)}. 
\label{eq_Bexpan}
\end{eqnarray}
By substituting Eqs. (\ref{eq_Aexpan}) and (\ref{eq_Bexpan}) into Eqs. 
(\ref{eq_A}) and (\ref{eq_B}), and linearizing the resulting equations in the
perturbation amplitudes $\hat{A}$ and $\hat{B}$, we find
\begin{eqnarray}
&&\partial_t \hat{A}(q_y,q_z,x,t) = \left [ \epsilon - \left (2iq_0
  \partial_x -q_y^2-q_z^2 \right )^2 - 2|A_0|^2 - 2|B_0|^2 \right ]
\hat{A}(q_y,q_z,x,t) \nonumber\\
&&\hskip 1.8cm -A_0^2 \hat{A}^*(-q_y,-q_z,x,t) -2A_0 B_0^* \hat{B}(q_y,q_z,x,t)
-2A_0B_0 \hat{B}^*(-q_y,-q_z,x,t), \label{eq_A_hat} \\
&&\partial_t \hat{B}(q_y,q_z,x,t) = \left [ \epsilon - \left (
  \partial_x^2 -q_y^2-2q_0q_z \right )^2 - 2|A_0|^2 - 2|B_0|^2 \right ]
\hat{B}(q_y,q_z,x,t) \nonumber\\
&&\hskip 1.8cm -B_0^2 \hat{B}^*(-q_y,-q_z,x,t) -2A_0^*B_0 \hat{A}(q_y,q_z,x,t)
-2A_0B_0 \hat{A}^*(-q_y,-q_z,x,t), \label{eq_B_hat}
\end{eqnarray}
where ``$*$'' denotes complex conjugation. 

Since the solution to the base state equations (\ref{eq_A0}) and (\ref{eq_B0})
cannot be obtained analytically, we proceed as follows: We consider an initial 
configuration that has a pair of symmetric grain boundaries so that periodic 
boundary conditions can be used in a numerical solution. Planar grain
boundaries on the $yz$ plane are located at $x=L_x/4$ and $3L_x/4$, where
$L_x$ is the system size along $x$ direction. The two boundaries need to be 
sufficiently far apart from each other so that their motion is independent. We 
have verified that this is the case for the system sizes used in our study.
We then numerically solve Eqs. (\ref{eq_A0}) and (\ref{eq_B0}), and use the
result to solve Eqs. (\ref{eq_A_hat}) and (\ref{eq_B_hat}) for given, fixed 
values of $q_y$ and $q_z$. A computational domain of size $L_x \times L_y
\times L_z$ is divided into an evenly spaced grid, with 8 grid points per 
wavelength of the base solution $\lambda_0=2\pi/q_0$. Hence the
corresponding grid spacing is $\Delta x = \Delta y = \Delta z = \lambda_0/8$.
We use a pseudo-spectral method with a Crank-Nicholson time stepping scheme 
for the linear terms, and a second order Adams-Bashford explicit algorithm 
for the nonlinear terms. A relatively large time step $\Delta t$ can be used 
while maintaining stability; $\Delta t=0.2$ has been used in all of our
calculations.

The stability of the base planar boundary is carried out indirectly by
introducing small random perturbations into  $\hat{A}$ and $\hat{B}$, with
both real and imaginary parts, and solving the initial value problem defined
by Eqs. (\ref{eq_A_hat}) and (\ref{eq_B_hat}). If the planar grain
boundary is stable, the perturbations $\hat{A}$ and $\hat{B}$
would decay in time for all wavenumbers $(q_y,q_z)$. However, our calculations 
show there exists a range of $(q_y,q_z)$ within which perturbations grow with 
time, indicating instability.

A typical result within the unstable region of wavenumbers is shown in
Fig. \ref{fig_AB_hat}. We consider $L_x=1024$ and $\epsilon=0.04$. A random
initial condition for the fields $\hat{A}$ and $\hat{B}$ has been considered, 
of zero average and uniformly distributed in $(-0.05,0.05)$. Far from the 
boundary region, the perturbations in $\hat{A}$ and $\hat{B}$ 
are seen to decay as a function of time, an observation which is consistent
with the fact that both bulk regions are in the stable region of the model's
stability diagram. On the contrary, both real and imaginary 
parts of the perturbations near the boundary grow in time for the example
chosen here ($q_y=47/128$ and $q_z=5/64$). The growth of the perturbation can be
quantified by writing $\hat{A}=|\hat{A}|e^{i\phi_A}$ and 
$\hat{B}=|\hat{B}|e^{i\phi_B}$, and as expected we find that
$|\hat{A}|$, $|\hat{B}|$ $\propto e^{\sigma t}$ with
$\sigma=\sigma(q_y,q_z)$ the perturbation growth rate.
We also note that the phase perturbation for $A$ becomes linear in space in
the unstable region: $\phi_A \propto -\delta q \cdot x$, whereas
$\phi_B$ remains uniform.

\begin{figure}
\centerline{\epsfig{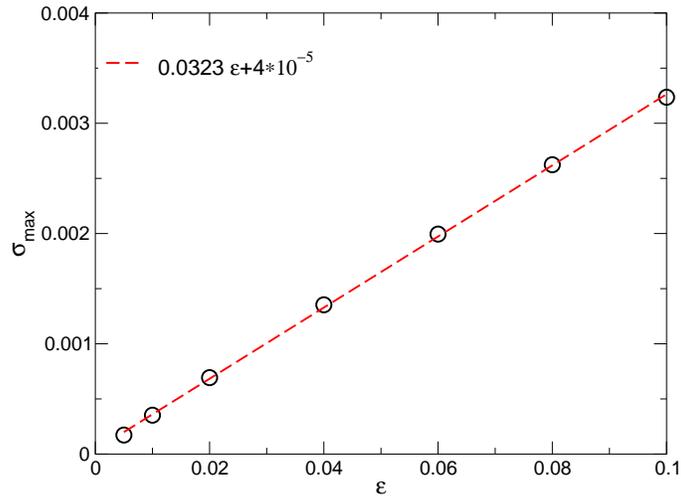}}
\caption{Maximum perturbation growth rate as a function of $\epsilon$,
  with system size $L_x=1024$. The dashed line is a linear fit to the data
  yielding a slope of $0.0323 \pm 0.0003$, and an intercept
  $(4 \pm 1) \times 10^{-5}$.
}
\label{fig_sigma_eps}
\end{figure}

We have repeated this process for a range of perturbation wavenumbers $q_{y}$
and $q_{z}$ and computed the growth rate $\sigma$. Our results are shown in 
Fig. \ref{fig_sigma} for $\epsilon=0.04$. We observe four unstable regions,
with maxima of the growth rate $\sigma^{\rm max}$ located at 
$(\pm q^{\rm max}_{y}, \pm q^{\rm max}_{z})$. As further shown in Fig. 
\ref{fig_sigma_eps}, $\sigma_{\rm max} > 0 $ for $\epsilon>0$, and linearly
increases with $\epsilon$. The characteristic wavenumbers for instability 
correspond to $q^{\rm max}_{y}$ and $q^{\rm max}_{z}$, and are plotted in
Fig. \ref{fig_Qeps} as filled circles ($q^{\rm max}_{y}$), and squares 
($q^{\rm max}_{z}$). The figure also shows the results of a direct numerical
solution of the Swift-Hohenberg equation (\ref{eq_s-h}) (pluses and
stars in Fig. \ref{fig_Qeps}), which we discuss below. Although it is
difficult to be certain about the limiting behavior as 
$\epsilon \rightarrow 0$ (the development of the instability is very slow
and the numerical results not very accurate), it appears from both
Figs. \ref{fig_sigma_eps} and \ref{fig_Qeps} that $\sigma^{\rm max}$, 
$q^{\rm max}_{y}$ and $q^{\rm max}_{z}$ tend to zero in that limit.

Further insight into the instability can be gained by direct numerical
solution of Eq. (\ref{eq_s-h}). The equation  
has been discretized on a evenly spaced grid of $256^3$
nodes, and integrated with the same method described above. 
A typical configuration following the boundary
instability is shown in Fig. \ref{fig_3d}. In this example,
$\epsilon=0.08$, and an initial configuration comprising two planar and
symmetric grain boundaries of wavenumber $q_{0}$ has been perturbed by a
random field, uniformly distributed between $(-0.05,0.05)$. In this case we
show in gray scale the value of the field $\psi$ at $t=2500$. The instability
manifests itself by finite wavenumber undulations of the grain boundary 
along both spatial directions, a perturbation that is of course precluded in
two dimensions.

We have determined the characteristic wavelengths of the instability
from the power spectrum of the order parameter $\psi$. Figure \ref{fig_psiq}
shows the structure factors $|\psi_{q_y}|$ (circles) and $|\psi_{q_z}|$
(squares), defined respectively as the one dimensional Fourier transform of 
$\psi$ on the boundary plane $x=L_x/4$ at fixed $z=L_z/2$, or fixed
$y=L_y/2$. Both power spectra exhibit peaks at nonzero wavenumbers along 
their respective direction. In the case shown in the figure,
$q^{\rm max}_{y}=0.4375$ and $q^{\rm max}_{z}=0.125$, so that 
$q^{\rm max}_{z} < q^{\rm max}_{y} < q_{0} = 1$. This result holds
for the entire range of $\epsilon$ that we have studied, and agrees with the
result of the stability analysis shown in Fig. \ref{fig_Qeps}. Note also that
the wavenumbers of maximum growth as determined from the power spectrum of
$\psi$ agree well with the values of $q^{\rm max}_{y}$ and 
$q^{\rm max}_{z}$ obtained from the stability analysis.

\section{Cross roll interaction in three dimensions}
\label{sec:cross-roll}

It may appear somewhat surprising that, given that the planar grain boundary 
configuration is uniform in the $y$ direction, and that the boundary is stable
against perturbations in the $xz$ plane, it should be unstable in three
dimensions as shown in Sec. \ref{sec:stability}. The projection to a two
dimensional lamellar pattern on the $xz$ plane is a 90$^{\circ}$ tilt 
grain boundary, which is known to be stable as long as both lamellar regions
have a wavenumber equal to $q_0$
\cite{re:manneville83b,re:tesauro87,re:malomed90,re:cross93,re:manneville90}. 
On the other hand, a two dimensional projection on the $xy$ plane 
involves coexistence of a region of lamellae A with a uniform
region B with constant value of $\psi$ (Fig. \ref{fig_3d}). This two
dimensional projection is clearly unstable, and boundary motion (from stable A
to unstable B) would occur driven by free energy reduction. However, any such
free energy difference between regions A and B is absent in three dimensions, and
therefore the origin of the instability deserves further scrutiny.

We begin the analysis by assuming the following approximate functional
dependence of the amplitudes $A$ and $B$ in a two dimensional system
\cite{re:zf2_jasnow},
\begin{equation}
A(x,z,t) \simeq A_0 (x-X(z,t)), \qquad B(x,z,t) \simeq B_0 (x-X(z,t)),
\label{eq_AB_int}
\end{equation}
under weak local distortion $X(z,t)$ of the grain boundary ($A_{0}$ and
$B_{0}$ are the stationary solutions given by Eqs. (\ref{eq_A0}) and
(\ref{eq_B0})). An equation of motion for the interface can be derived by
substituting Eqs. (\ref{eq_AB_int}) into the amplitude equations (\ref{eq_A}) 
and (\ref{eq_B}) (without the terms involving $\partial_y^2$), and expanding
to first order in $X$. By using $X(z,t)=\sum_{q_z} \hat{X}_q (t) e^{i q_z z}$, 
we find that $\hat{X}_q (t) = \hat{X}_q (0) e^{\sigma_X(q_z) t}$, with
$\sigma_X$ the growth rate of the perturbation in $X$ given by
\begin{eqnarray}
\sigma_X(q_z) &=& - \frac{1}{\int_{-\infty}^{\infty} dx
  \left ( |\partial_x A_0|^2 + |\partial_x B_0|^2 \right )} \nonumber\\
&\times& \int\limits_{-\infty}^{\infty} dx \left [
  |\partial_x A_0|^2 \left ( -\epsilon + |A_0|^2 + 2|B_0|^2 + q_z^4 
  \right ) + |\partial_x B_0|^2 \left ( -\epsilon + 2|A_0|^2 + |B_0|^2 
  + 4q_0^2 q_z^2 \right ) \right. \nonumber\\
&&\left. +4q_0^2 |\partial_x^2 A_0|^2 + |\partial_x^3 B_0|^2
  +(\partial_x |A_0|^2)^2/2 +(\partial_x |B_0|^2)^2/2
  +2(\partial_x |A_0|^2)(\partial_x |B_0|^2) \right ].
\label{eq_sigma_X}
\end{eqnarray}
We note from Eq. (\ref{eq_sigma_X}) both that $\sigma_X(q_z=0)=0$, and that 
$\sigma_X$ decreases monotonically with $q_z^2$. Therefore $\sigma_X<0$ 
for all nonzero $q_z$. Within this approximation, a perturbed grain boundary 
always relaxes to the stationary planar state, as is already well known.

Consider next the analog of Eq. (\ref{eq_AB_int}) for a three
dimensional system, 
\begin{equation}
A(x,y,z,t) \simeq A_0 (x-X(y,z,t)), \qquad B(x,y,z,t) \simeq B_0 (x-X(y,z,t)).
\end{equation}
The resulting interfacial equation to first order in $X$ is similar to Eq. 
(\ref{eq_sigma_X}), with some extra terms proportional to $q_y^2$ and
$q_y^4$. It is easy to show that the same results follows, namely
$\sigma_X(q_y,q_z) < 0$ for all $(q_y,q_z)$, i.e., a stable grain boundary, 
contrary to the calculations shown in Sec. \ref{sec:stability}.

\begin{figure}
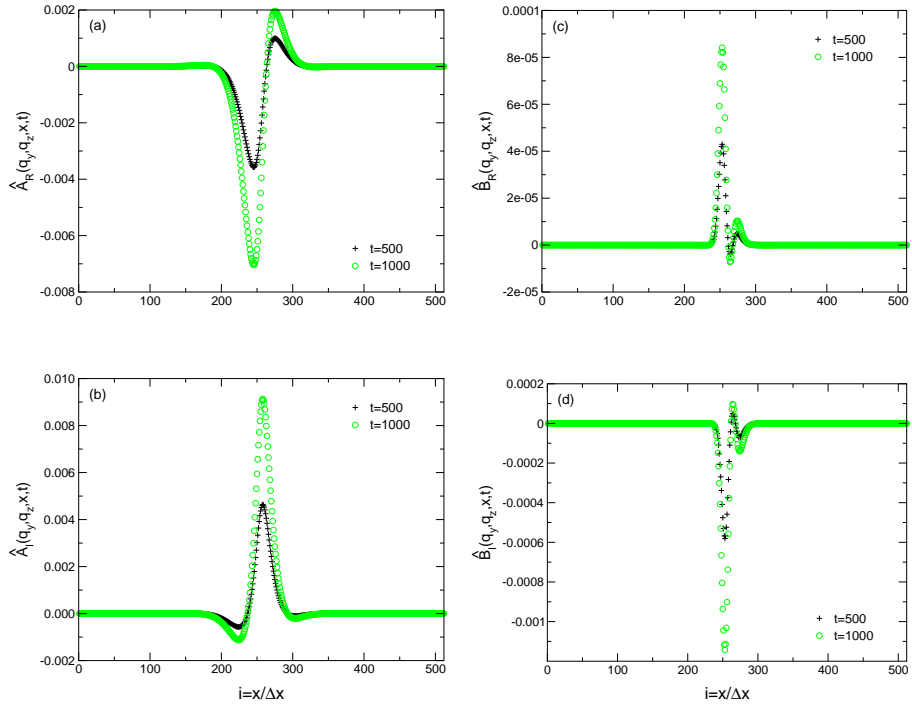

\centerline{
  \epsfig{figure=fig2a.eps,width=2.3in} \hskip 0.2cm
  \epsfig{figure=fig2c.eps,width=2.3in}}
\vskip 0.85cm
\centerline{
  \epsfig{figure=fig2b.eps,width=2.3in} \hskip 0.2cm
  \epsfig{figure=fig2d.eps,width=2.3in}}
\caption{Perturbation amplitudes $\hat{A}$ and $\hat{B}$
  as a function of grid index $i$ along the $x$ direction, 
  for wavenumbers $q_y=47/128$ and $q_z=5/64$, and
  $L_x=1024$, $\epsilon=0.04$, and initial noise amplitude $0.05$.
  Both real ($\hat{A}_R$ and $\hat{B}_R$) and imaginary ($\hat{A}_I$
  and $\hat{B}_I$) parts are shown at times $t=500$
  (pluses) and $1000$ (circles).
}
\label{fig_AB_hat}
\end{figure}

\begin{figure}
\centerline{\epsfig{figure=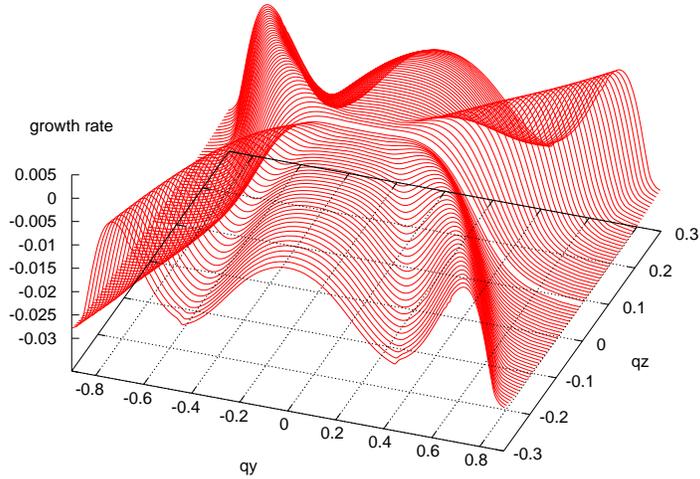,width=4.5in}}
\caption{Perturbation growth rate $\sigma$ as a function of
  wavenumbers $q_y$ and $q_z$, with the same values of the parameters $\epsilon$,
  $L_x$, and the initial noise amplitude as those of Fig. \ref{fig_AB_hat}.
  The maximum growth rate is found to be $1.35\times 10^{-3}$,
  corresponding to $4$ symmetric wavenumber positions 
  $(\pm q_y^{\rm max}, \pm q_z^{\rm max})$, with $q_y^{\rm
  max}=47/128$ and $q_z^{\rm max}=5/64$.
}
\label{fig_sigma}
\end{figure}

The inconsistency in the three dimensional results can be traced back to the
fact that assumption (\ref{eq_AB_int}) does not allow for an imaginary part of 
$A$ or $B$ because the stationary solutions $A_0$ and $B_0$ are both real.
Therefore no phase winding is allowed, a result which is consistent with
wavenumber selection in two dimensions, but that does not seem to be observed
in three dimensions (cf. Figs. \ref{fig_AB_hat}b and \ref{fig_AB_hat}d). 
We therefore argue that growth of phase perturbations is one of the
major causes of the instability in three dimensions.

In order to illustrate how combined phase and amplitude modulations around the
grain boundary can lead to a decrease in free energy, we recall that the
amplitude equations (\ref{eq_A}) and (\ref{eq_B}) can be written in gradient form
\begin{equation}
\partial_t A = -\delta {\cal F} / \delta A^*, \qquad
\partial_t B = -\delta {\cal F} / \delta B^*,
\nonumber
\end{equation}
with a potential ${\cal F}$
\begin{eqnarray}
{\cal F} = \int\int\int dx dy dz  &\biglb [&
  -\epsilon \left ( |A|^2 + |B|^2 \right ) 
  + \left ( |A|^4 + |B|^4 \right )/2 + 2 |A|^2 |B|^2  \nonumber\\
&+&\left | (2iq_0 \partial_x + \partial_y^2 + \partial_z^2)
  A \right |^2 + \left | (\partial_x^2 + \partial_y^2 +2iq_0
  \partial_z) B \right |^2 \bigrb ].
\label{eq_F}
\end{eqnarray}
The net change $\Delta {\cal F}$ relative to the planar grain boundary is
\begin{equation}
\Delta {\cal F} = {\cal F} - {\cal F}_0,
\label{eq_dF}
\end{equation}
where ${\cal F}_0$ of the base state is obtained by replacing $A$ and
$B$ in Eq. (\ref{eq_F}) with the stationary solutions $A_0(x)$ and
$B_0(x)$. By using the Fourier expansions (\ref{eq_Aexpan}) and
(\ref{eq_Bexpan}), we have calculated $\Delta {\cal F}$ up to second
order in the perturbations $\hat{A},\hat{B}$: $\Delta {\cal F} =
\Delta {\cal F}^{(1)} + \Delta {\cal F}^{(2)}$. The first order
result for $\Delta {\cal F}^{(1)}$ is
\begin{eqnarray}
\Delta {\cal F}^{(1)} = V_{yz} \int dx &\biglb [&
  2\left ( -\epsilon + |A_0|^2 + 2|B_0|^2 \right ) Re(A_0 \hat{A}^*(0))
  \nonumber\\
&+&2\left ( -\epsilon + 2|A_0|^2 + |B_0|^2 \right ) Re(B_0 \hat{B}^*(0))
  \nonumber\\
&+&8q_0^2 Re(\partial_x A_0 \ \partial_x \hat{A}^*(0))
  + 2 Re(\partial_x^2 B_0 \ \partial_x^2 \hat{B}^*(0)) \bigrb ],
\label{eq_dF1}
\end{eqnarray}
with $V_{yz}=L_y \times L_z$, $\hat{A}(0)=\hat{A}(q_y=q_z=0,x,t)$, and
$\hat{B}(0)=\hat{B}(q_y=q_z=0,x,t)$. Given the numerical solution for the
amplitudes of the perturbation described in Sec. \ref{sec:stability}, we find
that $\Delta {\cal F}^{(1)}$ is negligible (except for an initial transient
related the fast local relaxation of the configuration in response to the
random field which is added to the initial configuration). Therefore, $\Delta
{\cal F}$ is mostly determined by the second order result
\begin{eqnarray}
\Delta {\cal F}^{(2)} &=& V_{yz} \int dx \sum\limits_{q_y,q_z} \Biglb \{
  \left ( -\epsilon + |A_0|^2 + 2|B_0|^2 \right ) |\hat{A}({\bf q})|^2 
  +\left ( -\epsilon + 2|A_0|^2 + |B_0|^2 \right ) |\hat{B}({\bf q})|^2
  \nonumber\\
&+& \left | (2iq_0 \partial_x - q_y^2 -q_z^2)\hat{A}({\bf q}) \right |^2
  +\left | (\partial_x^2 -q_y^2 -2q_0 q_z)\hat{B}({\bf q}) \right |^2
  \nonumber\\
&+& \frac{1}{2} \left [ \left | A_0^* \hat{A}({\bf q}) 
    + A_0 \hat{A}^*(-{\bf q})\right |^2
    +\left |B_0^* \hat{B}(-{\bf q}) + B_0 \hat{B}^*({\bf q})\right |^2 
    \right ] \nonumber\\
&+& 2 \left [ A_0^*\hat{A}({\bf q}) + A_0 \hat{A}^*(-{\bf q}) \right ]
  \left [ B_0^* \hat{B}(-{\bf q}) + B_0 \hat{B}^*({\bf q}) \right ]
  \Biglb \}, \label{eq_dF2}
\end{eqnarray}
with $\hat{A}({\bf q})=\hat{A}(q_y,q_z,x,t)$, and $\hat{B}({\bf q})
=\hat{B}(q_y,q_z,x,t)$. $\Delta {\cal F}<0$ indicates instability against the
perturbation, while $\Delta
{\cal F}>0$ refers to the energy penalty of any modulations with the
system relaxing to its stationary base state.

\begin{figure}
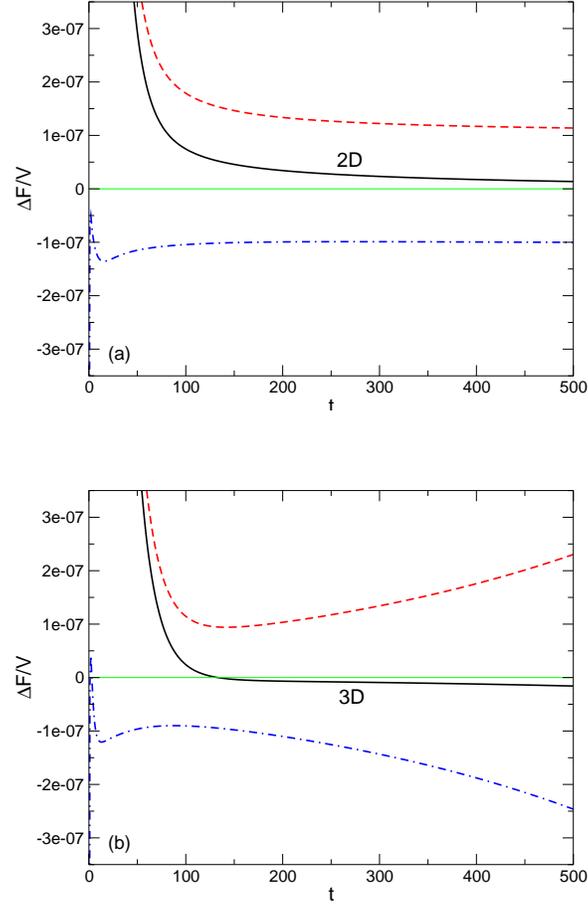

\vskip 1cm
\centerline{\epsfig{figure=fig7a.eps,width=3.in}}
\vskip 0.95cm
\centerline{\epsfig{figure=fig7b.eps,width=3.in}}
\caption{Time evolution of effective free energy density (per unit
  volume $V=L_x \times L_z$ or $V=L_x \times L_y \times L_z$) of
  the perturbed state, for (a) two and (b) three dimensional systems. In both 
  (a) and
  (b), the net energy change $\Delta {\cal F}$ (thick solid curve) is the
  combination of the negative contribution from the last term of
  Eq. (\ref{eq_dF2}) (dash dotted curve) and all the remaining (positive)
  terms (dashed curve). The parameters used here are the same as
  in Fig. \ref{fig_AB_hat}, except that in the two dimensional system 
  $|q_z|=1/128$ has been used in the calculation.
}
\label{fig_F}
\end{figure}

We now estimate $\Delta {\cal F}$ in both two and three dimensions by 
approximating the sums in Eq. (\ref{eq_dF2}) by the values at 
$(\pm q^{\rm max}_{y}, \pm q^{\rm max}_{z})$, the wavenumbers associated with
the largest growth rate $\sigma_{\rm max}$, and present the results in 
Fig. \ref{fig_F}. In two dimensions, $\Delta 
{\cal F}$ (given by Eq. (\ref{eq_dF2}) with $q_y=0$) remains positive for all 
times as expected (solid line). There is a negative contribution to 
$\Delta {\cal F}$ arising from the last term of Eq. (\ref{eq_dF2}) which 
reflects the coupling between $A$ and $B$ modes (dot-dashed line in Fig.
\ref{fig_F}a). This term only contributes to the energy integral in the 
boundary region, and is negligible otherwise. However, the remaining positive
contributions to Eq. (\ref{eq_dF2}) dominate (dashed line in the figure),
leading to the overall stability of the boundary. On the other hand,
the cross mode coupling in three dimensions dominates over the remaining 
stabilizing terms, leading to overall instability, as shown in
Fig. \ref{fig_F}b. It is therefore the additional phase space available for
the coupling between $A$ and $B$ modes at the boundary that is responsible for 
the instability in three dimensions.

\section{Discussion and conclusions}
\label{sec:discussion}

We have found a new instability associated with a $90^{\circ}$ tilt grain
boundary in a three dimensional lamellar phase, both from direct numerical 
solution of the Swift-Hohenberg model equation, and an analysis of the
corresponding amplitude equation. The latter result is therefore more generic
and applies to grain boundaries in phases of smectic symmetry.
The mode of instability is anisotropic on the grain boundary plane,
with wavelengths along both directions that are larger than the wavelength of
the base lamellar pattern. Of the two, the larger wavelength is directed along 
the normal to the lamellae oriented normal to the boundary plane. Both the
characteristic wavenumbers of the instability and the growth rate increase
with $\epsilon$. 

The instability is accompanied by phase perturbation, a mode that is absent in 
two dimensions, and that results from the cross coupling of the two lamellar 
modes in the boundary region. This coupling is also present in two dimensions, 
but it is not strong enough to produce instability.

\begin{figure}
\centerline{\epsfig{figure=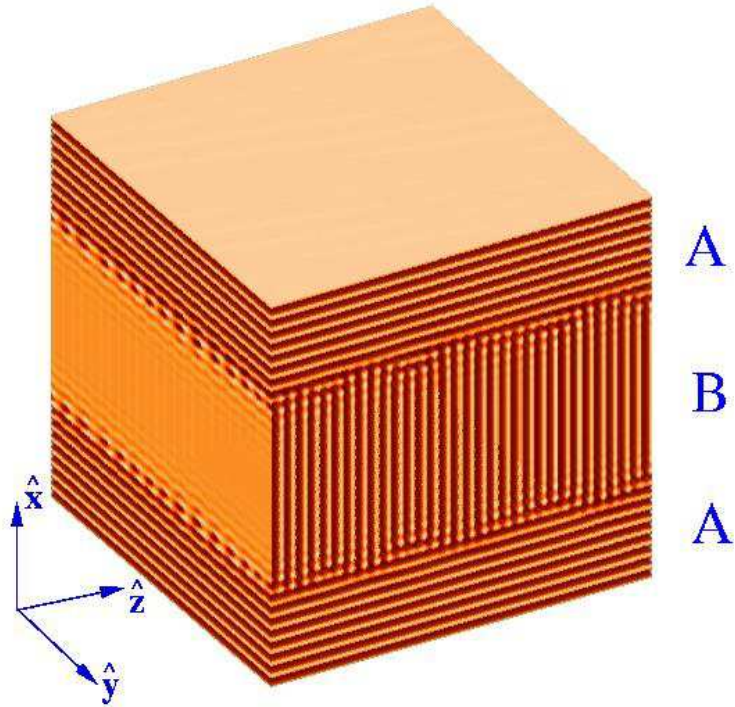,width=4.in}}
\caption{Grain boundary configuration obtained from numerical integration
  of the Swift-Hohenberg equation (\ref{eq_s-h}) in a $256^3$ system,
  for $\epsilon=0.08$, and an initial configuration comprising two symmetric
  grain boundaries as discussed in the text, plus small random noise uniformly
  distributed
  between $(-0.05,0.05)$. At the time shown ($t=2500$), the
  instability is readily apparent as undulations along both $y$ and $z$
  axes.
}
\label{fig_3d}
\end{figure}

\begin{figure}
\centerline{\epsfig{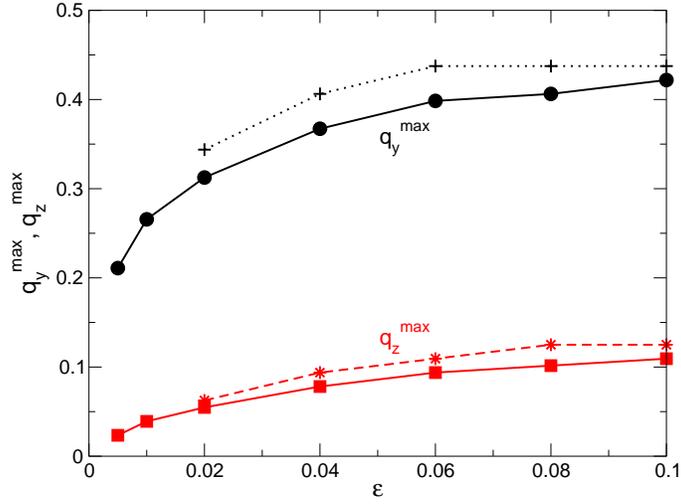}}
\caption{Characteristic wavenumbers of instability $q_y^{\rm max}$ and 
  $q_z^{\rm max}$ along two orthogonal
  directions of the grain boundary as a function of
  $\epsilon$. Symbols (+ and $\ast$) correspond to results obtained from direct
  numerical solution of the model equation (\ref{eq_s-h}) with system 
  size $256^3$, while filled circles and squares with solid
  lines have been determined from the stability analysis of the amplitude
  equations (\ref{eq_A}) and (\ref{eq_B}) with $L_x=1024$.
}
\label{fig_Qeps}
\end{figure}

\begin{figure}
\centerline{\epsfig{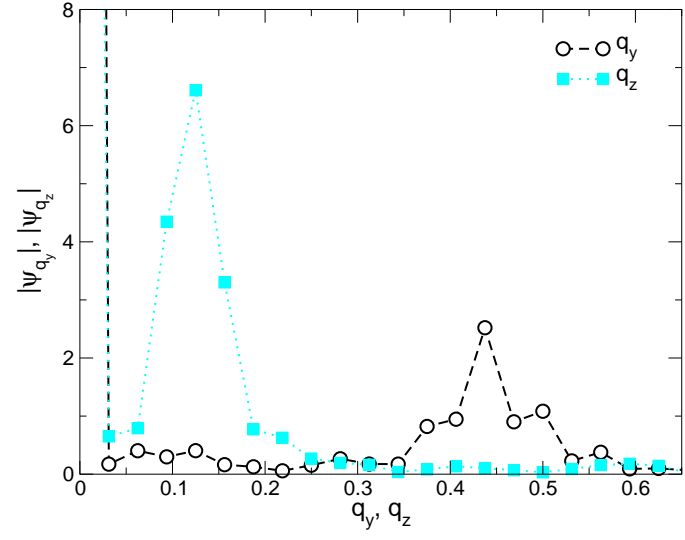}}
\caption{One-dimensional structure factor $|\psi_{q_y}|$ (at
  $z=L_z/2$) or $|\psi_{q_z}|$ (at $y=L_y/2$) on the grain boundary plane
  $x=L_x/4$ as a function of wavenumber $q_y$ or $q_z$ respectively. The
  parameters are the same as those of Fig. \ref{fig_3d}, except for
  $t=1650$.
}
\label{fig_psiq}
\end{figure}

It is important to verify that this instability is not an artifact of
finite size effects inherent to all our calculations. It is known, for
example, that there is net boundary attraction
when their separation is not much larger than their width, the latter
scaling as $\epsilon^{-1/2}$ for small $\epsilon$ \cite{re:manneville83b}. 
Our system sizes have been chosen large enough so that the pair of grain
boundaries in the computational domain remain stationary in a two dimensional 
geometry. Spot checks of the results shown in
Figs. \ref{fig_3d}, \ref{fig_Qeps}, and \ref{fig_psiq}, with different 
system sizes have revealed no discernible change.

The instability described above is different from that which is caused by 
deviations of the base lamellar wavenumber $q_x$ (in region A) or $q_z$ (in
region B) from $q_0$ \cite{re:manneville83b,re:tesauro87}. Such an instability
is also present in three dimensions. We have studied grain
boundary configurations in three dimensions with different values of
$q_x=q_z=q_0+\Delta q$, with $\Delta q$ small. Our calculations show that the
boundaries move toward each other (from region A to B) as
long as $\Delta q \neq 0$. The wavenumber $q_x$ of lamellae A always
relaxes to the optimum value $q_0=1$, leading to the creation of new lamellae
for $\Delta q >0$ (with $\Delta q$ small enough to remain inside the
stability region of lamellar phase \cite{re:greenside84}), or to the
disappearance of existing lamellae for $\Delta q <0$,
similar to the two dimensional results of Ref. \onlinecite{re:tesauro87}. During
the process, the wavenumber $q_z$ of lamellae B remains constant, with the 
extent of region B decreasing with time due to the invasion by
lamellae A. In the case of $\Delta q <0$, lamellae B also undergo a zig-zag 
instability.

In summary, our results indicate that $90^\circ$ tilt grain boundaries should be
rare in three dimensional extended samples, but readily observable in two
dimensional systems. This is in agreement with experimental findings in three
dimensional lamellar phases \cite{re:gido94}, and with numerical evidence in
the case of two dimensions \cite{re:cross93,re:manneville90}.
Following the boundary instability in three dimensions, we observe
growth and a limited amount of coarsening near the boundary region. However, 
the evolution is very slow, and we cannot determine whether a stationary state
will be reached, or whether evolution will continue until the two boundaries
annihilate each other.

\begin{acknowledgments}
This research has been supported by the National Science Foundation 
under grant DMR-0100903, and by NSERC Canada.
\end{acknowledgments}

\bibliographystyle{prsty}

\begin{thebibliography}{10}
\bibitem{re:zf2_jasnow}
We are indebted to David Jasnow for discussions about interfacial equations of
  motion.
\end{thebibliography}

\newpage

\end{document}